\documentclass[12pt,a4paper]{article}
\usepackage{amsfonts,latexsym}
\usepackage{graphicx,color}
\usepackage{dcolumn}
\usepackage{graphicx}
\usepackage{amsmath}
\usepackage{amsfonts}
\usepackage{amssymb}

\renewcommand{\thefootnote}{\fnsymbol{footnote}}

\begin{document}

\vspace{12mm}

\begin{center}
{{{\Large {\bf Conditions for superradiant instability of  the Kerr-Newman black holes  }}}}\\[10mm]

Yun Soo Myung$^{a,b}$\footnote{e-mail address: ysmyung@inje.ac.kr}\\[8mm]

{${}^a$Institute of Basic Sciences and Department  of Computer Simulation, Inje University Gimhae 50834, Korea\\[0pt]}

{${}^b$Asia Pacific Center for Theoretical Physics, Pohang 37673, Korea}

\end{center}
\vspace{2mm}

\begin{abstract}
We find two conditions for  superradiant instability of Kerr-Newman black holes under a charged massive scalar perturbation by analyzing asymptotic scalar potential and
far-region wave function. Actually, they  correspond to the condition  for getting a trapping well. Also, we obtain the conditions for superradiant stability of Kerr-Newman black holes which states that there is no trapping well.  The  analysis is applied to  Kerr black  holes to find a condition for superradiant instability.

\end{abstract}
\vspace{5mm}

\vspace{1.5cm}

\hspace{11.5cm}
\newpage
\renewcommand{\thefootnote}{\arabic{footnote}}
\setcounter{footnote}{0}


\section{Introduction}
If a light boson exist with proper mass,  gravitational bound states are formed  around rotating charged black holes. These bound states could extract continuously
electromagnetic  or rotational energy from black holes~\cite{Arvanitaki:2010sy}. It is a  superradiance phenomenon in the black holes~\cite{Brito:2015oca}.
The existence of superradiant modes can be converted into an instability of the black hole background if a mechanism of mirror  to trap these modes is installed  near the black hole. This is called superradiant instability. If a scalar  has a mass $\mu$, its mass would act as a reflecting mirror~~\cite{Damour:1976kh}.
The superradiant instability of the Kerr black hole was found for $M\mu\gg1$~\cite{Zouros:1979iw}, $M\mu\ll1$~\cite{Detweiler:1980uk}, and   $M\mu\le0.5$~\cite{Dolan:2007mj}.   A scalar potential  including  a trapping well is essential for generating  a quasibound state~\cite{Zouros:1979iw,Arvanitaki:2010sy,Konoplya:2011qq} whose wave function ia peaked far outside the ergo region. If there is no trapping well in the potential, it may correspond to the superradiant stability with bound state.

In the Kerr-Newman  black hole (KNBH) with mass $M$, charge $Q$, and angular momentum $J$, the superradiant instability condition for a charged massive scalar with mass $\mu$ and charge $q$  was firstly obtained as $qQ<\mu M$  which may be   a  condition for  a trapping well~\cite{Furuhashi:2004jk}. However, $qQ<\mu M$ is not satisfied simultaneously if one imposes the superradiance condition ($\omega<\omega_c$ with $\omega_c=m\Omega_H+q\Phi_H$) and thus, it is  a condition for bound states ~\cite{Herdeiro:2013pia}. In other words, the condition of $qQ<\mu M$ corresponds to the Newton-Coulomb requirement for the gravitational force to exceed the electrostatic force.  Also, it is noted  that  their effective potential $V_{\rm eff}(r)$ seems  not to be   a correct form.
Scalar clouds with $\omega=\omega_c$  and $\omega<\mu$ were obtained  in~\cite{Hod:2014baa,Benone:2014ssa}. The absorption cross section  was recently found  to give  a negative absorption cross section for co-rotating spherical waves~\cite{Benone:2019all}.

On the other hand, there was an approach  to analyzing   superradiant instability based on the scalar potentials~\cite{Huang:2016qnk}.
We stress  that  the appearance/disappearance  of a trapping well is a decisive  condition for superradiant instability/stability.
Recently, it was shown that the superradiant stability of KNBHs under a charged massive scalar perturbation can be
achieved if $qQ>\mu M$ and $r_-/r_+\le 1/3$ are satisfied~\cite{Xu:2020fgq}, in addition to $\omega<\omega_c$ and the bound state condition ($\omega<\mu$).  However, their potential based on the analysis seems to be incorrect because  $\Psi_{\ell m}=\sqrt{\Delta}R_{\ell m}$ is used  and a  tortoise coordinate $r_*$ defined by $dr_*=(r^2+a^2)dr/\Delta$ is not used by following Ref.~\cite{Hod:2012zza}. So, it suggests that  $qQ>\mu M$ is not  a condition for superradiant stability.
The superradiant stability of a charged massive scalar  based on a desirable potential  was discussed in the KNBH background~\cite{Myung:2022kex}.

In this work, we wish to find two conditions for getting a trapping well of KNBH under a charged massive scalar perturbation by analyzing asymptotic scalar potential $V_{aaKN}(r)$ and
far-region wave function $U[p,s;cr]$. They are given by  $V'_{aaKN}(r)>0~(M\mu^2>qQ\omega)$ and $U'[p,s;cr]>0~(p<0)$. In addition, the
conditions for no trapping well are  $V'_{aaKN}(r)<0~(M\mu^2<qQ\omega)$ and $U'[p,s;cr]<0~(p>0)$.
We apply  the same analysis to Kerr black hole under a massive scalar propagation to find the condition for a trapping well.

\section{Potentials around KNBHs}
Let us  introduce the
Boyer-Lindquist coordinates to represent a  KNBH  with mass $M$, charge $Q$,  and angular momentum $J$
\begin{eqnarray}
ds^2_{\rm KN}&=&\bar{g}_{\mu\nu}dx^\mu dx^\nu \nonumber \\
&=&-\frac{\Delta}{\rho^2}\Big(dt -a \sin^2\theta d\phi\Big)^2 +\frac{\rho^2}{\Delta} dr^2+
\rho^2d\theta^2 +\frac{\sin^2\theta}{\rho^2}\Big[(r^2+a^2)d\phi -adt\Big]^2 \label{KN}
\end{eqnarray}
with
\begin{eqnarray}
\Delta=r^2-2Mr+a^2+Q^2,~ \rho^2=r^2+a^2 \cos^2\theta,~{\rm and}~a=\frac{J}{M}.
 \label{mps}
\end{eqnarray}
We choose  the electromagnetic potential as
\begin{equation}
\bar{A}_\mu=\frac{Q r}{\rho^2}\Big(-1,0,0, a\sin^2\theta\Big).
\end{equation}
The outer and inner horizons are obtained from  $\Delta=(r-r_+)(r-r_-)=0$ ($\bar{g}^{rr}=0$) as
\begin{equation}
r_{\pm}=M\pm \sqrt{M^2-a^2-Q^2}.
\end{equation}
One describes a charged massive scalar perturbation $\Phi$  on the background of KNBHs by adapting the perturbed scalar equation
\begin{equation}
(\bar{\nabla}^\mu-i q \bar{A}^\mu)(\bar{\nabla}_\mu-i q \bar{A}_\mu)^*\Phi-\mu^2\Phi=0.\label{phi-eq1}
\end{equation}
Considering  the static and  axis-symmetric
background (\ref{KN}), it is plausible  to separate the scalar perturbation
into modes
\begin{equation}
\Phi(t,r,\theta,\phi)=\Sigma_{lm}e^{-i\omega t + i m \phi} S_{l m
}(\theta) R_{l m}(r)\,, \label{sep}
\end{equation}
where $S_{\ell m}(\theta)$ is  spheroidal harmonics with $-m\le \ell
\le m$ and $R_{l m}(r)$ describes  a radial part of the wave
function. Substituting  (\ref{sep}) into (\ref{phi-eq1}), we have  the
angular equation  for $S_{l m}(\theta)$ and the  Teukolsky equation as~\cite{Hod:2014baa}
\begin{eqnarray}
&& \frac{1}{\sin \theta}\partial_{\theta}\Big(
\sin \theta
\partial_{\theta} S_{\ell m}(\theta) \Big )+ \left [\lambda_{lm}+ a^2 (\mu^2-\omega^2) \sin^2
{\theta}-\frac{m^2}{\sin ^2{\theta}} \right ]S_{l m}(\theta) =0,
\label{wave-ang1}
\end{eqnarray}
\begin{eqnarray}
\Delta \partial_r \Big( \Delta \partial_r R_{\ell m}(r) \Big)+U(r)R_{lm}(r)=0,
\label{wave-rad}
\end{eqnarray}
where
\begin{eqnarray}
U(r)=[\omega(r^2+a^2)-am-qQr]^2+\Delta[2am\omega-\mu^2(r^2+a^2) -\lambda_{lm}]. \label{u-pot}
\end{eqnarray}
We note that Eq.(\ref{wave-rad}) could be used directly for computing absorption cross section, quasinormal modes of the scalar, and scalar clouds.
At this stage, we introduce  the tortoise  coordinate $r_*$  defined  by $dr_*=
\frac{r^2+a^2}{\Delta}dr$ to derive the Schr\"odinger-type equation.
In this case, an interesting region of $r\in[r_+,\infty)$ could be mapped into the whole region of  $r_*\in(-\infty,\infty)$.
Then,  the radial equation (\ref{wave-rad}) takes
the Schr\"odinger-type equation when choosing $\Psi_{lm}=\sqrt{a^2+r^2} R_{lm}$
\begin{equation}
\frac{d^2\Psi_{lm}(r_*)}{dr_*^2}+\Big[\omega^2-V_{KN}(r)\Big]\Psi_{lm}(r_*)=0, \label{sch-eq}
\end{equation}
where  the potential $V_{KN}(r)$ is found to be~\cite{Benone:2014ssa}
\begin{eqnarray}
V_{KN}(r)=\omega^2&-&\frac{3\Delta^2r^2}{(a^2+r^2)^4}+\frac{\Delta[\Delta+2r(r-M)]}{(a^2+r^2)^3} \nonumber \\
&+&\frac{\Delta \mu^2}{a^2+r^2}-\Big[\omega-\frac{am}{a^2+r^2}-\frac{qQr}{a^2+r^2}\Big]^2\nonumber \\
&-&\frac{\Delta}{(a^2+r^2)^2}\Big[2am\omega -\lambda_{lm}\Big] \label{c-pot}.
\end{eqnarray}
Replacing  $\lambda_{lm}$ by $\tilde{\lambda}_{lm}+a^2(\omega^2-\mu^2)$ with $\tilde{\lambda}_{lm}=l(l+1)+\cdots$, one finds a familiar  angular equation
\begin{eqnarray}
&& \frac{1}{\sin \theta}\partial_{\theta}\Big(
\sin \theta
\partial_{\theta} S_{\ell m}(\theta) \Big )+ \left [\tilde{\lambda}_{lm}+ a^2 (\omega^2-\mu^2) \cos^2
{\theta}-\frac{m^2}{\sin ^2{\theta}} \right ]S_{l m}(\theta) =0.
\label{wave-ang2}
\end{eqnarray}
Before we proceed, we would like to mention  a superradiant scattering of the scalar of the KNBHs.
We find two limits such that $V_{KN}(r\to \infty)\to \mu^2$ and $V_{KN}(r\to r_+) \to \omega^2 -(\omega-\omega_c)^2$.
In this case, we have scattering forms of plane waves as~\cite{Brito:2015oca}
\begin{eqnarray}
\Psi_{lm}&\sim&  e^{-i\sqrt{\omega^2-\mu^2} r_*}(\leftarrow)+{\cal R}e^{+i\sqrt{\omega^2-\mu^2} r_*}(\rightarrow),\quad r_*\to +\infty(r\to \infty) , \label{asymp1}\\
\Psi_{lm}&\sim& {\cal T} e^{-i(\omega-\omega_c) r_*}(\leftarrow),\quad r_*\to -\infty(r\to r_+) \label{asymp2}
\end{eqnarray}
with the ${\cal T}({\cal R})$ the transmission (reflection) amplitudes. A  wrongskian $W(\Psi,\Psi^*) $ condition  of
$i \frac{d}{dr_*} W(\Psi,\Psi^*)=0$  leads to
\begin{equation}
|{\cal R}|^2=1-\frac{\omega-\omega_c}{\sqrt{\omega^2-\mu^2}}|{\cal T}|^2,
\end{equation}
which implies that outgoing waves with $\omega>\mu$ propagate to infinity and the superradiant scattering occurs ($|{\cal R}|^2>|{\cal I}|^2$) whenever the superradiance condition is satisfied 
\begin{equation}
\omega<\omega_c. \label{s-cond}
\end{equation}
Curiously, the superradiance  is associated to  having  a negative absorption cross section~\cite{Benone:2019all}. For a KNBH, the total absorption cross section becomes negative for co-rotating spherical waves at low frequencies.

Now, we wish  to describe the superradiant instability briefly.
The two boundary conditions imply an exponentially decaying wave (bound state) away from trapping well and a purely outgoing wave (superradiance) near the outer horizon under Eq.(\ref{s-cond}):
\begin{eqnarray}
\Psi&\sim& e^{-\sqrt{\mu^2-\omega^2}r},\quad r_* \to \infty(r\to \infty), \label{rad-sol1} \\
\Psi&\sim& e^{-i(\omega-\omega_c)r_*},\quad r_* \to -\infty(r\to r_+).  \label{rad-sol2}
\end{eqnarray}
From Eq.(\ref{rad-sol1}), one needs  the bound state condition to obtain an exponentially  decaying mode
\begin{equation}
 \omega<\mu. \label{b-cond}
\end{equation}
Furthermore, the  superradiant instability/stability  is determined  by a shape of the potential, in addition to Eqs.(\ref{s-cond}) and (\ref{b-cond}).
Importantly, a key condition for superradiant instability  is to include  a positive trapping well in the potential.
If there is no trapping well in the potential, it implies  a superradiant stability.

To find out the condition  of a  trapping well, we have to  consider  a  potential $V_{aKN}(r)$ obtained when expanding $V_{KN}(r)$  in the far-region
\begin{equation}
V_{aKN}(r)=\mu^2-\frac{2(M \mu^2-q Q \omega)}{r}+\frac{\lambda_{lm}+Q^2(\mu^2-q^2)}{r^2}. \label{aKN}
\end{equation}
Its first derivative
\begin{equation}
V'_{aKN}(r)=\frac{2(M \mu^2-q Q \omega)}{r^2}-\frac{2[\lambda_{lm}+Q^2(\mu^2-q^2)]}{r^3}, \label{faKN}
\end{equation}
where the mass term $\mu^2$ in Eq.(\ref{aKN}) is missed. We note that there is no restrictions on parameters in deriving $V_{aKN}(r)$.
Here, one tempts to say that the condition for  trapping well (no trapping well)  is given by $V'_{aKN}(r)>0~(V'_{aKN}(r)<0)$.
Nevertheless, it is difficult to find any analytic condition from $V'_{aKN}(r)>0~(V'_{aKN}(r)<0)$.
To this end, we consider an asymptotic potential $V_{aaKN}(r)$ obtained when expanding $V_{KN}(r)$ at $r=\infty$  and its first derivative
\begin{equation}
V_{aaKN}(r)=\mu^2-\frac{2(M \mu^2-q Q \omega)}{r},\quad V'_{aaKN}(r)=\frac{2(M \mu^2-q Q \omega)}{r^2}. \label{aa-pot}
\end{equation}
Here,  one requires $V'_{aaKN}(r)>0$ ($M\mu^2>qQ\omega$)  for arising a trapping well, whereas  $V'_{aaKN}(r)>0~(M\mu^2<qQ\omega) $ is demanded  for no trapping well.
 However, this is not a sufficient condition for getting a trapping well.
In the next section, we will find  the other condition by analyzing far-region scalar function. The other condition for  trapping well (no trapping well) will be given by $U'[p,s;cr]>0~(U'[p,s;cr]<0)$ where $U[p,s,cr]$ is the confluent hypergeometric function.

At this stage, we display two potentials, their far-region and asymptotic potentials where one includes a trapping well with $M\mu^2>qQ\omega$ [(Left) Fig. 1] and the other does not include a trapping well but it matches up with $M\mu^2>qQ\omega$  [(Right) Fig. 1].
Fig. 2 indicates a potential with a trapping well for $q>\mu$ and $M\mu^2>qQ\omega$, showing a feature of a charged massive scalar propagating around the KNBH.
It is worth noting that all have $V'_{aaKN}(r)>0$. One needs to find the other condition to have no trapping well.
\begin{figure*}[t!]
   \centering
  \includegraphics{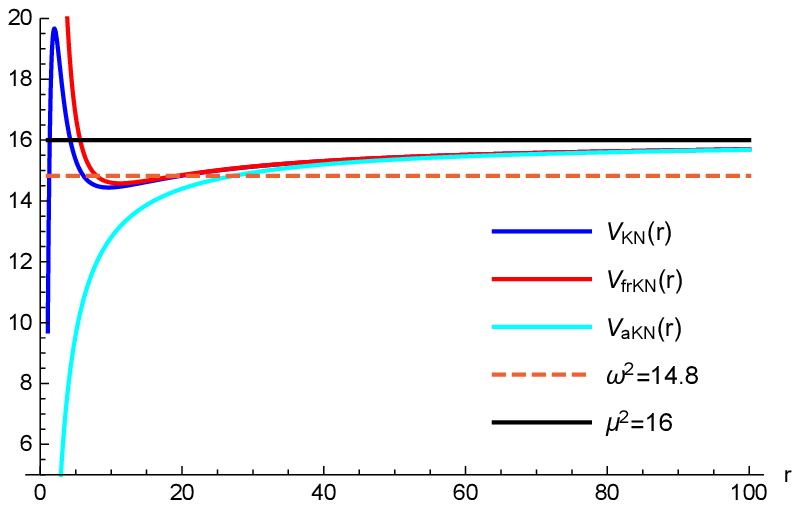}
  \hfill%
  \includegraphics{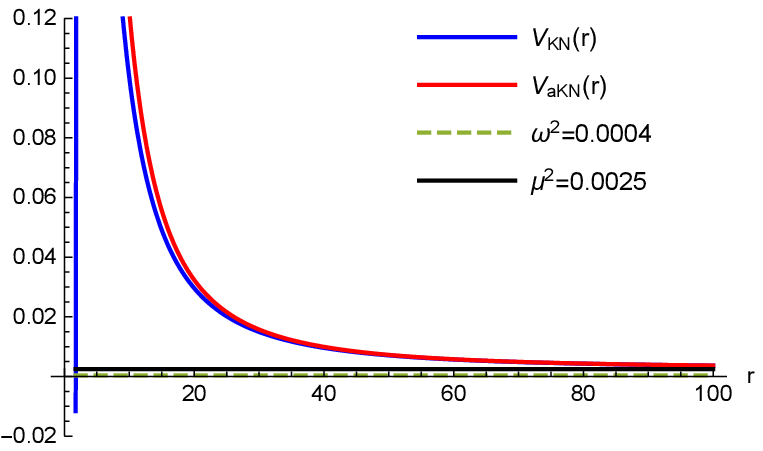}
\caption{(Left) Potential with trapping well $V_{KN}(r)$, its far-region $V_{aKN}(r)$, and  asymptotic potential $V_{aaKN}(r)$ as functions of $r\in[r_+=1.06,100]$ with $M=1,Q=0.01,\omega=3.85,a=0.998,m=13,q=0.2,\lambda_{lm}=180,\mu=4$.
$V_{KN}(r)$ has a trapping well located at $r=9.61$.
We check   $\omega<\omega_c(=6.11)$ and $\omega<\mu$ as two  conditions for superradiant instability. (Right) Potential  without trapping well $V_{KN}(r)$ as function of $r\in[r_+=1.741,100]$ with $M=1,Q=0.6,\omega=0.02,a=0.3,m=1,q=0.1,\lambda_{lm}=12,\mu=0.05$. We check two conditions of  $\omega<\omega_c(=0.13)$ and $\omega<\mu$.  }
\end{figure*}
\begin{figure*}[t!]
   \centering
  \includegraphics{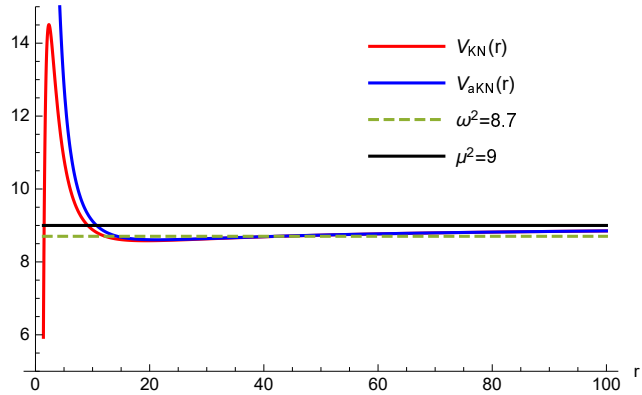}
\caption{(Left) Potential with trapping well $V_{KN}(r)$, its far-region $V_{aKN}(r)$, and  asymptotic potential $V_{aaKN}(r)$ as functions of $r\in[r_+=1.39,100]$ with $M=1,Q=0.01,\omega=2.95,a=0.9,m=13,q=20,\lambda_{lm}=180,\mu=3$. Here, we have $q>\mu$, showing a feature of KNBH.
$V_{KN}(r)$  has a trapping well located at $r=19.3$.
We check that  two conditions of    $\omega<\omega_c(=4.38)$ and $\omega<\mu$  are satisfied  for superradiant instability.   }
\end{figure*}

Finally, we describe   two conditions for superradiant instability and stability under a charged massive scalar propagating around the KNBHs:\\
(i) superradiant instability $\to$ $\omega<\omega_c$ and  $\omega<\mu$  with  a positive trapping well.\\
(ii) superradiant stability $\to$ $\omega<\omega_c$ and $\omega<\mu$  without a positive trapping well. \\

\section{Condition for trapping well}

One needs to observe  far-region scalar functions   to distinguish between trapping well and no trapping well.
For this purpose, we wish to derive a scalar equation  in the far-region.

In the far-region where taking  $r_*\sim r$ approximately, we obtain an  equation from Eqs. (\ref{sch-eq}) with (\ref{aKN}) as
\begin{equation}
\Big[\frac{d^2}{dr^2}+\omega^2-V_{aKN}(r)\Big]\Psi_{lm}(r)=0. \label{asymp-eq}
\end{equation}
The above equation  could be rewritten as
\begin{equation}
\Big[\frac{d^2}{dr^2}-A^2+\frac{B}{r}-\frac{C}{r^2}\Big]\Psi_{lm}(r)=0, \label{ABC-eq}
\end{equation}
where three coefficients are given by
\begin{equation}
A=\sqrt{\mu^2-\omega^2},\quad B=2(M\mu^2-qQ\omega),\quad C=\lambda_{lm}+Q^2(\mu^2-q^2).
\end{equation}
The solution is given exactly  by the Whittaker function $W[p,s;cr]$   and the confluent  hypergeometric function $U[p,s;c r]$ as
\begin{eqnarray}
\Psi_{lm}(r)&=&c_1 W\Big[\frac{B}{2A},k;2A r\Big] \label{wavef-0} \\
&=&c_1 e^{-Ar} (2A r)^{k+\frac{1}{2}}
~U\Big[k+\frac{1}{2}-\frac{B}{2A},1+2k;2A r\Big], \label{wavef-1}
\end{eqnarray}
where 
\begin{equation}
k=\frac{1}{2}\sqrt{1+4C}.
\end{equation}
Here, we find a bound state of  $e^{-\sqrt{\mu^2-\omega^2}r}$ appeared  in (\ref{rad-sol1}) and $M\mu^2>qQ\omega~(M\mu^2<qQ\omega)$ corresponding to $B>0~(B<0)$. In the asymptotic region, one has an asymptotic wave function
\begin{equation}
\Psi^{A}_{lm}(r)\sim e^{-\sqrt{\mu^2-\omega^2}r} \Big(2 \sqrt{\mu^2-\omega^2} r\Big)^{\frac{M\mu^2-qQ\omega}{\sqrt{\mu^2-\omega^2}}},
\end{equation}
which always leads to zero as $r\to\infty$ for $\sqrt{\mu^2-\omega^2}>0$.

Let us observe a radial mode  $\Psi_{lm}(r)$ for  a trapping well [see (Left) Fig. 1], implying superradiant instability.
As is shown in (Left) Fig. 3, Eq.(\ref{wavef-1}) shows a  quasibound state whose wave function is a peak located far from the outer horizon.
A similar picture is found from (Left) Fig. 4 which is based on Fig. 2 with $q>\mu$. This case indicates  a feature of a charged massive scalar propagating around the KNBH.

Contrastively, we consider a radial mode for a potential  without trapping well [shown in (Right) Fig. 1 with $V'_{aaKN}>0(M\mu^2>qQ\omega)$], implying superradiant stability.
As is shown in Fig. 5, Eq.(\ref{wavef-1})   shows  an exponentially decaying mode. This case explains why one needs to find the other condition for no trapping well.

\begin{figure*}[t!]
   \centering
  \includegraphics{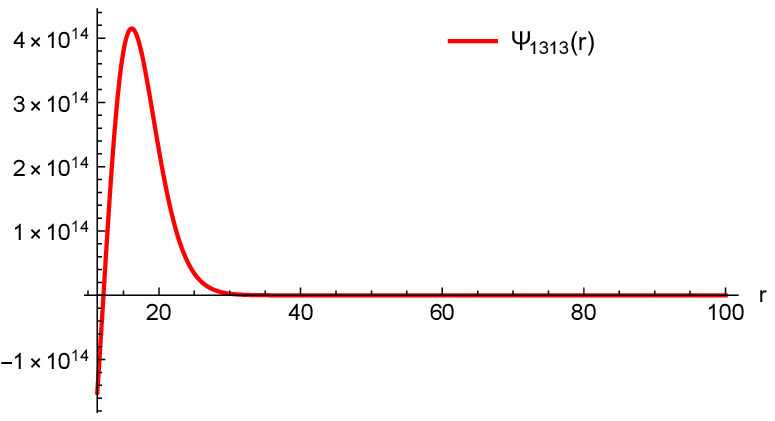}
  \hfill%
  \includegraphics{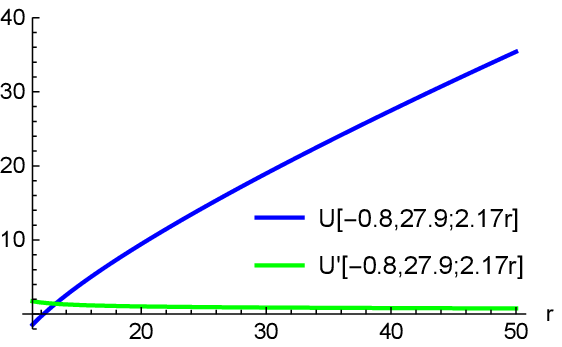}
\caption{ (Left) Radial mode showing quasibound state $\Psi_{1313}(r)$ as a function of $r\in[11.3,100]$ with trapping well. (Right) Confluent hypergeometric function $U[-0.8,27.8;2.17 r]$ is an increasing function of $r$ and its derivative $U'[-0.8,27.8;2.17 r]$ is positive.  All parameters go together with (Left) Fig. 1.  }
\end{figure*}
\begin{figure*}[t!]
   \centering
  \includegraphics{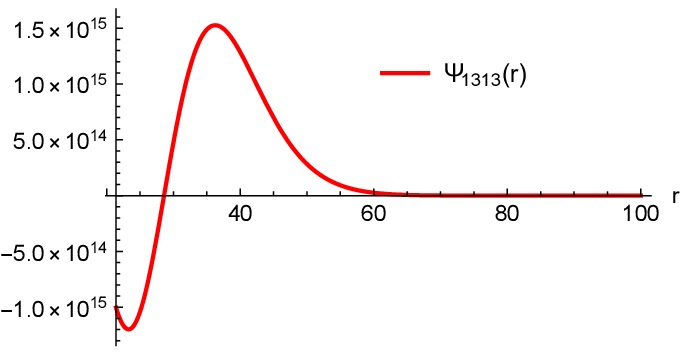}
  \hfill%
  \includegraphics{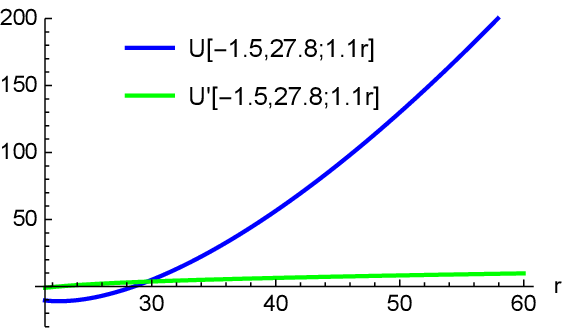}
\caption{ (Left) Radial mode showing quasibound state $\Psi_{1313}(r)$ as a function of $r\in[21.4,100]$ with trapping well. (Right) Confluent hypergeometric function $U[-1.5,27.8;1.1 r]$ is an increasing function of $r$ and its derivative $U'[-1.5,27.8;1.1 r]$ is positive.  All parameters go together with  Fig. 2.  }
\end{figure*}
\begin{figure*}[t!]
   \centering
  \includegraphics{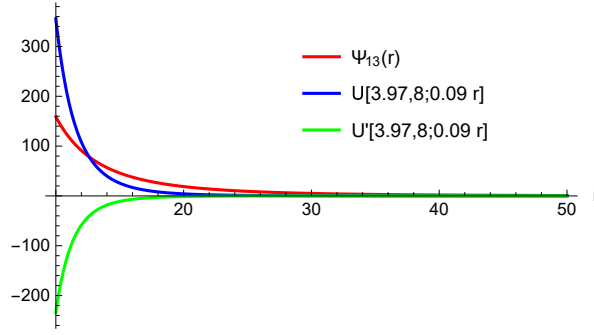}
\caption{ Bound state function $\Psi_{13}(r)$  as $r\in[5,100]$ without trapping well. Confluent hypergeometric function $U[3.97,8;0.09r]$ represents  a decreasing function of $r$ and its derivative $U'[3.97,8;0.09r]$ is  negative. All parameters go together with (Right) Fig. 1. }
\end{figure*}
At this stage, it is useful to introduce the asymptotic form of $U[p,s;cr]$ as~\cite{Myung:2022gdb}
\begin{equation}
U[p,s;cr\to \infty]\rightarrow\quad  (cr)^{-p}\Big[1-\frac{p(1+p-s)}{cr} +{\cal O}\Big(\frac{1}{cr}\Big)^2\Big]. \label{asymp-h}
\end{equation}
Here, one observes an increasing function $U[p,s;cr]$ for a negative $p$ [(Right) Fig. 3 and (Right) Fig. 4], while one finds  a decreasing function  for a positive $p$ (Fig. 5).
Furthermore, considering the first derivative of $U[p,s;cr]$ with respect to $r$ ($c>0$) as
\begin{equation}
U'[p,s;cr]=-pcU[1+p,1+s;cr], \label{d-hU}
\end{equation}
it implies that  the condition for a trapping well is
\begin{equation}
U'[p,s;cr]>0 \to \quad p<0, \label{dp-hU}
\end{equation}
whereas the condition for no trapping well is given by
\begin{equation}
U'[p,s;cr]<0 \to \quad p>0.\label{dn-hU}
\end{equation}
Therefore, the quasibound state with a peak  could be found when $p$  is negative as
\begin{equation}
p<0 \to \quad \sqrt{1+4C}<\frac{B}{A}-1 \to \quad \frac{M\mu^2-qQ\omega}{\sqrt{\mu^2-\omega^2}}>k+\frac{1}{2}, \label{trap-well}
\end{equation}
which is  the other condition for a trapping well, in addition to $M\mu^2>qQ\omega$ $(V'_{aaKN}(r)>0)$.
\begin{figure*}[t!]
   \centering
  \includegraphics{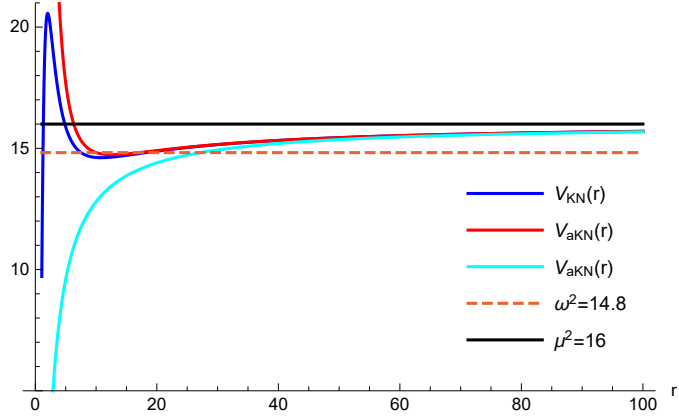}
\caption{ Potential with trapping well $V_{KN}(r)$,  its far-region $V_{aKN}(r)$, and  asymptotic potential $V_{aaKN}(r)$ as functions of $r\in[r_+=1.062,100]$ with $M=1,Q=0.01,\omega=3.85,a=0.998,m=13,q=0.2,\lambda_{lm},\mu=4$. Here, we choose $\lambda_{lm}=202.461$ to find  $a=-5.9\times 10^{-9}\sim 0$ in $U[p,s;cr]$.
$V_{KN}(r)$ has a trapping well located at $r=10.8$.}
\end{figure*}
\begin{figure*}[t!]
   \centering
  \includegraphics{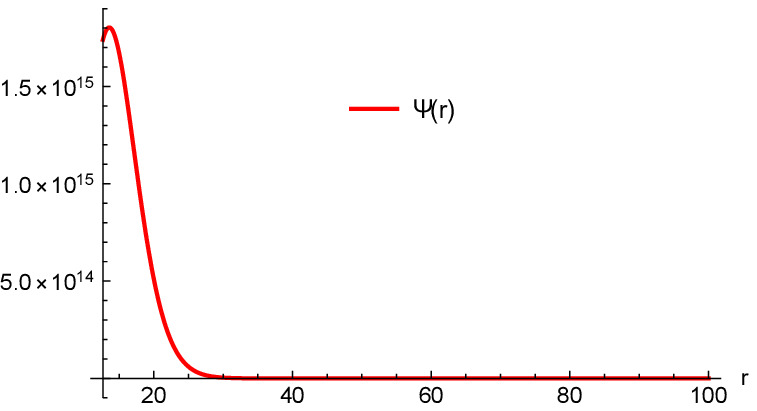}
  \hfill%
  \includegraphics{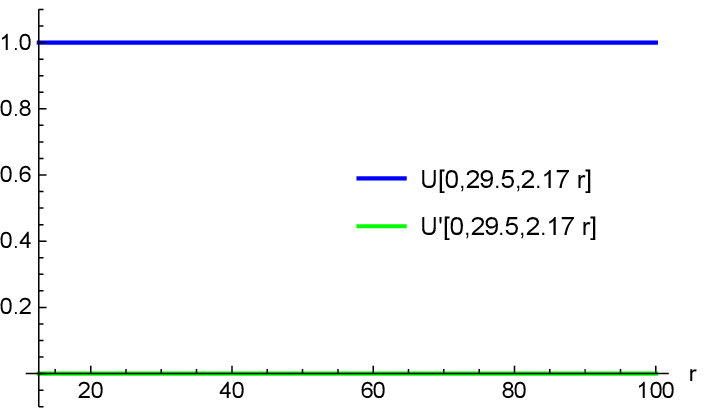}
\caption{ (Left) Radial mode $\Psi(r)$ showing  a half of a peak  as a function of $r\in[12.66,100]$ with trapping well. (Right) Confluent hypergeometric function $U[-5.9\times 10^{-9}\simeq 0,29.5;2.17 r]$  represents a constant and its derivative $U'[-5.9\times 10^{-9}\simeq 0,29.5;2.17 r]$ is nearly zero.  All coefficients go together with  Fig. 6.  }
\end{figure*}
 On the other hand,
the bound state   could be found  for a positive $p$ as
\begin{equation}
p>0  \to \quad \sqrt{1+4C}>\frac{B}{A}-1\to \quad \frac{M\mu^2-qQ\omega}{\sqrt{\mu^2-\omega^2}}<k+\frac{1}{2},\label{no-tw}
\end{equation}
which denotes  the other condition for no trapping well, in addition to $M\mu^2<qQ\omega$ $[(V'_{aaKN}(r)<0)]$.
We could not find  a condition of $M\mu/qQ<1$  for superradiant stability~\cite{Xu:2020fgq}. However, we  obtain  one condition from $V'_{aaKN}(r)<0$ as
\begin{equation}
\frac{M\mu}{qQ}<\frac{\omega}{\mu}<1. \label{one-cond}
\end{equation}

Finally, we consider  the case of $p\simeq0$ to denote a boundary state between quasibound and bound states. Here, we  display  a corresponding potential in  Fig. 6 which seems to have a trapping well.
The asymptotic wave function shows  a half of peak [(Left) Fig. 7] and thus, it does not represent  a quasibound state completely.  Its confluent hypergeometric function $U[p,29.5;2.17]$ with $p=-5.9\times 10^{-9}\simeq0$ is a nearly constant [(Right) Fig. 7], which implies  $U'[p,29.5;2.17]\simeq0$. This completes our classification: $p<0,~p=0,~p>0$.

\section{Condition for trapping well in Kerr black holes}
Now, we are in a position to study the superradiant instability/stability of Kerr black holes under a massive scalar propagation.
We obtain this case when choosing $q=Q=0$. Its far-region  equation is given by Eq.(\ref{ABC-eq}) with $A=\sqrt{\mu^2-\omega^2},~B=2M\mu^2,C=l(l+1),$ and $k=l+1/2$.
Here, one should find the other condition for no trapping well because  $B>0~[V'_{aaK}(r)>0]$ implies that a trapping well necessarily arises.
In this case, the solution is given   by~\cite{Guo:2021xao}
\begin{eqnarray}
\Psi^{K}_{lm}(r)=c_2 e^{-\sqrt{\mu^2-\omega^2}r} (2\sqrt{\mu^2-\omega^2} r)^{l+1}
~U\Big[l+1-\frac{M\mu^2}{\sqrt{\mu^2-\omega^2}},2l+2;2\sqrt{\mu^2-\omega^2} r\Big]. \label{wavef-K1}
\end{eqnarray}
Here, the condition for a trapping well is obtained from the first argument ($p<0$) in $U[p,s;cr]$ as
\begin{equation}
\frac{M\mu^2}{\sqrt{\mu^2-\omega^2}}>l+1, \label{K-si}
\end{equation}
while the condition for no trapping well  takes the form
\begin{equation}
\frac{M\mu^2}{\sqrt{\mu^2-\omega^2}}<l+1.\label{K-ss}
\end{equation}
 Eq.(\ref{K-si}) is satisfied  for the potential with trapping well [similar to (Left) Fig. 1] whose parameters are given by $M=1,~\omega=3.85,~a=0.998, m=13,~l=13,~\mu= 4,$ and $\omega_c=6.101>\omega$.
 We note that its confluent hypergeometric function $U[-0.61, 28, 2.19 r]$ is an increasing function of $r$.
 Also, we have checked the stability condition (\ref{K-ss}) for the potential without trapping well [similar to (Right) Fig. 1] whose parameters are $M=1,~\omega= 0.02,~a= 0.3, m=1,~l=3,~\mu=0.05$, and $\omega_c=0.076>\omega$. As is expected, its confluent hypergeometric function $U[3.95, 8, 0.09 r]$ is a decreasing function of $r$.

 For a complex $\omega=3.85+10^{-6}i~(|\omega_I|\ll \omega_R,~\omega_R<m\Omega_H)$ with the same parameters~\cite{Guo:2021xao,Kodama:2011zc}, one has its asymptotic  solution
 \begin{equation}
 \Psi_{13,13}\sim (51422 - 2.35 i) e^{(-1.09+3.5\times 10^{-6} i)r}r^{14}~U[-0.74-5\times 10^{-5}i,28,(2.17-7\times 10^{-6}i)r]
 \end{equation}
 whose real part takes the form of a peak as in (Left) Fig. 3. Here, one may rewrite $p$ in $U[p,s;cr]$ as
 \begin{equation}
 p=-n-\delta\nu,
 \end{equation}
where $n=1$ and $\delta\nu=-0.26+5\times 10^{-5} i(|\delta\nu|=0.26<1)$. The latter complex number might represent a deviation from the
hydrogen wave functions.  A state of superradiant instability could not be approximately described by a small shift of  hydrogen  energy level because $|\delta\nu|=0.26$ is not a small shift.  This may be so because we use $M\mu=4>1$ (not ultralight bosons with $\omega\sim \mu \ll 1/M$).

\section{Discussions}

First of all, we would like to mention the conditions for superradiant instability and stability.
The superradiant instability can be achieved for  $\omega<\omega_c$ and  $\omega<\mu$  with  a positive trapping well,
whereas the superradiant stability can be found for  $\omega<\omega_c$ and $\omega<\mu$  without a positive trapping well.
The presence of a trapping well is regarded as a decisive condition for superradiant instability. If there is no trapping well, it corresponds to the superradiant stability.

In this work, we have firstly obtained  two conditions for getting a trapping well of KNBHs under a charged massive scalar perturbation by analyzing asymptotic scalar potential [$V_{aaKN}(r)$] and far-region
wave function ($U[p,s;cr]$). They are given by  $V'_{aaKN}(r)>0~(M\mu^2>qQ\omega)$ and  $U'[p,s;cr]>0~(p<0)$. Also, the two
conditions for no trapping well are   $V'_{aaKN}(r)<0~(M\mu^2<qQ\omega)$ and   $U'[p,s;cr]<0~(p>0)$. From the former, we have derived one condition of Eq.(\ref{one-cond}) for the superradiant stability.

We have  carried out the same analysis for Kerr black hole under a massive scalar propagation to find the conditions for superradiant instability and stability.
The superradiant instability condition is given by $M\mu^2/\sqrt{\mu^2-\omega^2}>l+1$, while the stability condition is $M\mu^2/\sqrt{\mu^2-\omega^2}<l+1$ because $V'_{aaK}(r)>0~(M\mu^2>0)$ is always satisfied.

 \vspace{0.5cm}

{\bf Acknowledgments}
 \vspace{0.5cm}

This work was supported by a grant from Inje University for the Research in 2021 (20210040).

\end{document}